\documentstyle[preprint,eqsecnum,aps]{revtex}

\begin{document}
\draft

\title{\Large \bf
Colored Noise in Spatially-Extended Systems}
\author{\large
J. Garc\'{\i}a-Ojalvo$^{a,b}$
and J.M. Sancho$^{b}$}
\address{\mbox{ } \\
(a) Departament de F\'{\i}sica i Enginyeria
Nuclear,\\ Escola T\`{e}cnica Superior d'Enginyers Industrials
de Terrassa\\ Universitat Polit\`{e}cnica de Catalunya,
Colom 11, E-08222 Terrassa, Spain.\\
\mbox{ } \\(b) Departament d'Estructura i Constituents de la
Mat\`{e}ria, Facultat de F\'{\i}sica
\\ Universitat de Barcelona,
Diagonal 647, E-08028 Barcelona, Spain.}
\maketitle
\begin{abstract}
We study the effects of time and space correlations of an
external additive colored noise on the steady-state behavior
of a Time-Dependent Ginzburg-Landau model. Simulations show
the existence of nonequilibrium phase transitions controlled
by both the correlation time and length of the noise.
A Fokker-Planck equation and
the steady probability density of the process
are obtained by means of a theoretical approximation.
\end{abstract}
\pacs{}

\narrowtext

\section{Introduction}
\label{sec:intr}

\subsection{General aspects}

Langevin equations in spatially-extended systems are accepted
as a common reference frame in the study of those equilibrium
and nonequilibrium phenomena in which fluctuations play a
relevant role. This kind of equations have been used in the
study of critical phenomena \cite{hohenberg77}, phase
separation dynamics \cite{gunton83}, instabilities in liquid
crystals \cite{kai86} and bistability in chemical reactions
\cite{dewel85,horsthemke84}, among a large variety of systems.
Fluctuations can have an internal
(thermal) or external origin with respect to the system under
study. Internal fluctuations have been considered in critical
phenomena and phase separation, whereas external fluctuations
have been mostly studied in relation to liquid crystals and
chemical reactions. In these last cases internal fluctuations
are also present, although they are much less relevant than
the external ones.

Statistical Mechanics shows that internal fluctuations produce
phase transitions from an ordered to a disordered state as
their intensity increases. These phenomena are described by
means of a singular behavior of the relevant variable (order
parameter) in the vicinity of the transition point. In this
paper we will consider the situation in which fluctuations are
external (not thermal) and study the possible existence of
nonequilibrium transitions controlled by this sort of
fluctuations.

A prototype Langevin equation can be written in the general
form,
\begin{equation}
\label{eq:genlang}
\dot{\psi}\left(\vec{r},s\right)=
f\left(\psi\left(\vec{r},s\right)
,\nabla,\alpha\right) + g\left(\psi\left(\vec{r},s\right)
,\nabla\right) \eta\left(\vec{r},s\right).
\end{equation}

The deterministic force $f$ depends in general on the field
variable $\psi$, its spatial derivatives and a set of control
parameters $\alpha$. This force comes either from
reaction-diffusion terms or from a free-energy functional. The
spatial derivatives model the coupling of the field at one
given point with its value in the neighborhood. The existence
of such a coupling
implies that the Langevin equation (\ref{eq:genlang}) will not
be an ordinary but a {\em partial} stochastic differential
equation, whose rigorous mathematical study is nowadays under
active research.

The stochastic force represents the influence of the
surroundings, a heat bath, other internal degrees of freedom
or a stochastic external control parameter. It is usually
supposed to be proportional to a noise term $\eta$. The form
of $g$ depends on the kind of coupling between the noise and
the variable. When the noise is not coupled to the field, $g$
takes a constant value and the noise is said to be {\em
additive}. When some kind of coupling between the field and
the noise exists, the function $g$ depends on the field. This is the
so-called {\em multiplicative} noise.

On behalf of the Central Limit Theorem the random variable
$\eta$ can be supposed to be gaussian distributed with zero
mean. Its correlation at different points and instants of time
will in general be given by
\begin{equation}
\label{eq:gennoi}
\left< \;\eta(\vec{r},s) \;\eta(\vec{r}\,',s') \;\right> =
D\; h\left( \frac{\vec{r}-\vec{r}\,'}{l_0}; \frac{\mid 
s-s' \mid}{\sigma}
\right)
\end{equation}
where $D$ is the intensity of the noise, $\sigma$ is its {\em
correlation time} and $l_0$ its {\em correlation length}. When
there is no correlation neither in space nor in time
the function $h$ becomes the product of two delta functions:
it is the {\em
white}-noise case. In any other case the noise is said to be
{\em colored}.

A noise accounting for fluctuations of internal (thermal)
origin is supposed to be uncoupled to the system. Moreover one
assumes that it is white in space and time. This is so because
the noise represents many microscopic degrees of freedom which
evolve in spatial and temporal scales much shorter than those
of the relevant variables of the system. Hence internal noise
is usually modelled by additive white noise.

When the origin of the noise is external there may exist a
coupling between the system and the fluctuations. Besides, in
general there is no difference between the time and length
scales of the noise and the field and one has to take into
account the fact that this noise could have some structure
either in space or in time. Thus external noise can be colored
and/or multiplicative.

We are going to focus on the case of a potential system
affected by an external additive noise ($g=1$). When this
noise is white the steady probability distribution of the
field can be obtained {\em exactly} for suitable boundary
conditions. When it is colored this is not possible in
general, and only {\em approximate} techniques can be used.
Actually, difficulties here arise from a correlation of the
noise in {\em time}, in which case the stochastic process
defined by Eq. (\ref{eq:genlang}) is {\em non-markovian}.
Correlation in space does not present this type of
mathematical difficulties. 

Non-markovian stochastic processes in zero dimensions (no
gradient terms in Eq. (\ref{eq:genlang})) have been thoroughly
studied by now (see Ref. \cite{moss89}, vol. 1, for a review). Among
the interesting systems where colored-noise effects have been
studied one finds lasers \cite{lett88}, chemical
reactions \cite{horsthemke84} and liquid crystals
\cite{kai86}. Time correlations in the noise happen to induce
transitions in such systems \cite{horsthemke84}. Due to the
zero-dimensional character of these systems, these transitions
can only be characterized by means of qualitative changes in
the shape of the steady probability distribution of the
variable (for instance, a change in the number of its minima
and maxima). These effects have been observed in analog
\cite{sancho82b} and digital simulations \cite{sancho82a} and
have been explained analytically \cite{sancho82a,sancho89}.

Stochastic processes in $d>0$ dimensions affected by
space-and-time colored noises
have not been studied so extensively.
In this case, and due to the spatial character of the system,
an ergodicity-breaking effect permits a phase transition to be
characterized by means of a singular behavior of an order
parameter of the system, its relative fluctuations and its
statistical moments and correlations, such as it is done in
equilibrium phase transitions.

In this paper, our aim is to study the effects of the
correlation time and length of an additive noise in the
behavior of a two-dimensional stochastic model.

\subsection{The Time Dependent Ginzburg-Landau
model with colored noise}

We will study a model given by the following Langevin field
equation:
\begin{equation}
\label{eq:tdglor}
\frac{\partial\phi\left(\vec{r},s\right)}{\partial s} = \Gamma
\left( b\phi -u\phi^3+k\nabla^2\phi\right) + \Gamma^{1/2}
\eta\left(\vec{r},s
\right).
\end{equation}
In the study of critical phenomena \cite{hohenberg77} and
phase-separation dynamics \cite{gunton83}, the noise term has
been assumed to be gaussian and white, with zero mean and
correlation
\begin{equation}
\label{eq:white}
\left<\;\eta\left(\vec{r},s\right)
\eta\left(\vec{r}\,',s'\right)\;\right>=
2 D\;\delta \left(\vec{r}-\vec{r}\,'\right) \delta(s-s')
\end{equation}
with $D = k_B T$ (the so-called {\em intensity} of the noise).
In this case the model is known as the non-conserved {\em
Time-Dependent Ginzburg-Landau} model (model A in the notation
of Ref. \cite{hohenberg77}). Here a
fluctuation-dissipation relation ensures that the steady-state
probability distribution of $\phi$ is given by the Boltzmann
expression in terms of the following Ginzburg-Landau free
energy:
\begin{equation}
\label{eq:glfe}
F\left[\phi(\vec{r},s)\right]=\int \left\{-\frac{1}{2}b\phi^2
+\frac{1}{4}u\phi^4+\frac{1}{2}k\left(\nabla\phi\right)^2
\right\} d\vec{r}
\end{equation}
When $b>0$, this function has a double-well structure which
can be overridden by a high-intensity noise. Thus a phase
transition will be observed from an ordered towards a
disordered state when the intensity of the noise increases
beyond a critical value \cite{toral90}.

In this paper we are interested in the effects of a correlated
noise. Hence we will assume that the stochastic force in Eq.
(\ref{eq:tdglor}) is also gaussian distributed with zero mean
but with a correlation given by the general expresion of Eq.
(\ref{eq:gennoi}). Now there is no fluctuation-dissipation
relation, so that one cannot expect to reach an equilibrium
steady state as the one given by the free energy
(\ref{eq:glfe}).

In order to analyze the effects of the noise correlations in
the dynamics of the stochastic process governed by Eq.
(\ref{eq:tdglor}) we start by removing all unnecessary
parameters. This will be done by means of an
adimensionalization of the equation through the following
change of variables:
\begin{equation}
\label{eq:change}
\phi=\sqrt{\frac{b}{u}}\;\psi,\;\;\;\;\;s=\frac{1}{2\Gamma
b}\;t,
\;\;\;\;\;\vec{r}=\sqrt{\frac{k}{b}}\;\vec{x}
\end{equation}
Now the Langevin equation corresponding to
our model is:
\begin{equation}
\label{eq:tdgl}
\frac{\partial\psi(\vec{x},t)}{\partial t} = \frac{1}{2}
\left( \psi - \psi^3 + \nabla^2 \psi \right) + \xi(\vec{x},t)
\end{equation}
\begin{equation}
\label{eq:tdgl-cor}
\left< \;\xi(\vec{x},t) \xi(\vec{x}\,',t') \;\right> =
\varepsilon \;h \left( \frac{\vec{x}-\vec{x}\,'}{\lambda} ;
\frac{\mid t-t'\mid}{\tau}\right)
\end{equation}
Thus we now have only three dimensionless
parameters
\begin{equation}
\label{eq:dimpar}
\varepsilon=\frac{u}{2b^{2-d/2}k^{d/2}} \; D,\;\;\;\;\;
\tau=2\Gamma b\; \sigma, \;\;\;\;\; \lambda=\sqrt{\frac{b}{k}}
\;l_0.
\end{equation}

The rest of the paper is organized as follows.
In Sec. \ref{sec:sim} the simulation procedure and its results
are presented and in Sec. \ref{sec:spd} a theoretical
approximate approach
to the problem is developed. Finally some conclusions and
comments are stated.

\section{A numerical approach}
\label{sec:sim}

Due to the nonlinear character of Eq. (\ref{eq:tdgl})
an exact theoretical treatment of the problem cannot be done,
and therefore a numerical approach is required.
Simulations are performed on a CRAY Y-MP computer where
vectorization, but not parallelization, is used.
Eqs. (\ref{eq:tdgl}) and (\ref{eq:tdgl-cor}) have been
simulated on a regular two-dimensional lattice with $L\times
L$ square cells of size $\Delta x = 1$. In this space, the
Langevin equation has the form
\begin{equation}
\label{eq:dle}
\frac{\partial \psi_i}{\partial t} =
f_i\left(\vec{\psi}\right)+\xi_i(t)
\end{equation}
where the cells have been named with one index independently
of the dimension of the discrete space. And the correlation
(\ref{eq:tdgl-cor}) of the noise in this discrete space is:
\begin{equation}
\label{eq:discor}
\left< \xi_i(t)\; \xi_j(t') \right> = \varepsilon
h_{ij}\left(\frac{\mid t-t'\mid}{\tau}\right) 
\end{equation}
The force is given by
\begin{equation}
\label{eq:f4dfunc}
f_i\left(\vec{\psi}\right) = \frac{1}{2} \left(\psi_i -
\psi_i^3 + \left(\nabla^2 \psi\right)_i \right)
\end{equation}
where $\left(\nabla^2 \psi\right)_i= \nabla^2_{ik}
\psi_k$ and $\nabla^2_{ik}$ is a discretised version of the
laplacian operator \cite{abramowitz} (repeated indexes are
summed up).

The stochastic correlated force $\xi_i(t)$ is simulated
by means the following Langevin equation \cite{ojalvo92b}:
\begin{equation}
\label{eq:lambda}
\dot{\xi}_i(t) = -\frac{1}{\tau}\;(\delta_{ij}-\lambda^2 \nabla_{ij}^2)
\; \xi_j + \frac{1}{\tau}\; \mu_i(t)
\end{equation}
which is a generalization of the evolution equation of
an Ornstein-Uhlenbeck process \cite{sancho82a}.
$\mu_i(t)$ is a gaussian white noise with zero mean
and intensity equal to $\varepsilon$, which can be efficiently
generated by means of a vectorizable approximate algorithm
implementing a numerical inversion method \cite{toral93}. The
laplacian term ensures a correlation in space of order
$\lambda$ and $\tau$ is the correlation time. Fig. 1 shows the
spatial decay of the spherically averaged correlation function
of the noise for several values of $\lambda$. There one can
see that $\lambda$ is indeed a measure of the correlation
length of the stochastic force. A similar picture is obtained
if the correlation is plotted versus $t-t'$ \cite{ojalvo92b}.

Eq. (\ref{eq:lambda}) is linear, so that it can be simulated
{\em exactly} in Fourier space \cite{ojalvo92b} and
antitransformed to real space along with the integration of
Eq. (\ref{eq:dle}). This last numerical integration cannot be exact
(${\displaystyle f_i\left(\vec{\psi}\right)}$ is not linear),
so that a second order Runge-Kutta algorithm has been used.
This type of algorithm allows us to take a relatively
large integration time step ($\Delta t = 0.05$) with
no loss of stability.

The quantities which are computed and analyzed in the
simulation are \cite{binder88,ojalvo92a}:
\begin{itemize}
\item
The steady mean density of the absolute value of the field:
\begin{equation}
M_1 = \frac{\left< m \right> } {L^2}
\label{eq:m1}
\end{equation}
where $m=\mid \sum_{i,j} \psi_{ij} \mid$, and the average is
made over the time evolution of the process in the steady
state and over different realizations of the noise.
\item
The relative fluctuations of the field:
\begin{equation}
M_2 = \frac{\left < m ^2 \right
> - \left < m \right >^2} {L^2 \varepsilon}
\label{eq:m2}
\end{equation}
\item
The linear relaxation time of the process, defined as
\begin{equation}
\label{eq:tc}
\tau_R^{-1} = - \left.\frac{d}{dt} C(t) \right|_{t=0}
\end{equation}        
where the correlation function $C(t)$ is
\begin{equation}
\label{eq:tcf}
C(t)=\frac{\left<m(t_0)m(t_0+t)\right>_{t_0}-\left<m
\right>^2}{\left<m^2\right>-\left<m\right>^2}
\end{equation}
\end{itemize}

The system is made to evolve from an initially ordered state,
$\psi=1$, and we let it relax for an interval of time large
enough to be close to the steady state, before performing the
time averages. On the other hand, the number of samples in the
colectivity averages is between $20$ and $40$. Periodic
boundary conditions are considered in all cases and a
finite-size analysis is performed by considering five
different system sizes. It is worth noting that in this kind
of numerical analysis the space discretization $\Delta x$ is
an independent parameter affecting the steady behavior of the
system \cite{toral90}, but we have not explored this situation
and have instead kept $\Delta x$ fixed and equal to $1$.

In a previous work \cite{ojalvo92a} we reported results
obtained for the case of a noise correlated in time
($\tau \neq 0$) but uncorrelated in space
($\lambda=0$). In this case the algorithm for the generation
of the noise is simpler than the Fourier algorithm mentioned
above. Fig. 2(a) shows
clearly the effect of $\tau$ in the transition, as obtained in
the simulations. When $\tau=0$ (the standard white-noise
case) a transition towards disorder is found at a critical
noise intensity $\varepsilon_c=0.38$. This result is in
agreement with a study of the $\phi^4$ model in the case of
white noise \cite{toral90}. As $\tau$ increases the
critical intensity is shifted towards higher values (the
peak of the relative fluctuations indicating the transition
moves to the right). This means that $\tau$ somehow "softens"
the effect of the noise, whose intensity needs to be higher to
destroy the initial ordered state. Fig. 2(b) shows the phase
diagram of this system, where a curve in the
$(\varepsilon,\tau)$ plane divides regions where the system is
ordered and disordered. The sign of the slope of this curve is
another indication of the "softening" effect of the
correlation time of the noise.

Fig. 3 shows the effects of $\tau$ on the three quantities
defined above for a fixed value of $\varepsilon$ and still for
$\lambda=0$. Peaks for the relative fluctuations and the
relaxation time are obtained for approximately the same value
of $1/\tau$ and a decay of the mean value of the field shows
that there is a transition from an ordered to a
disordered state as $\tau$ decreases. We have thus found a phase
transition controlled by the correlation time of the noise.

In Fig. 4 results are presented for fixed values of
$\varepsilon$ and $\tau$ against a decreasing value of
$\lambda$. A similar behavior is obtained, showing the
existence of a phase transition controlled by the correlation
length of the noise. The role of the correlation length is
also a "softening" one.

A finite-size scaling analysis applied to the two transitions
found above allows us to evaluate the position of the critical
point in both cases and to give an estimation of the values of
the critical exponents associated to them. Concerning the
$\tau$-controlled transition (Fig. 3), the critical point is
found to be located at $\tau_c^{-1}=1.0$, whereas in the case
of the $\lambda$-controlled transition (Fig. 4) we find
$\lambda_c^{-1}=1.9$. These transition points are determined
by extrapolating to infinite size the position of the maximum of
both the relative fluctuations $M_2$ and the linear relaxation
time $\tau_R$, and the results obtained by the two methods coincide
within the estimated numerical error ($\sim 10 \%$). Static
critical exponents can be calculated by means of the following
finite-size scaling relations \cite{toral90,barber83}:
\begin{eqnarray}
\mid &\alpha&_c(L) - \alpha_c \mid \;\sim \; L^{-1/\nu}
\nonumber \\
&M&_1 \sim \;\mid \alpha_c - \alpha \mid^\beta
\nonumber \\
&M&_2^{max} \sim  L^{\gamma/\nu}
\end{eqnarray}
where $\alpha = \tau^{-1}$ or $\lambda^{-1}$.
The dynamic critical exponent z can be found with the
relation \cite{landau88}
\begin{equation}
\tau_R^{max} \sim L^z
\end{equation}
The numerical results obtained for these exponents
for the $\tau$-transition (Fig. 3) are $\beta \simeq 0.19$,
\linebreak $\nu \simeq 0.78$, $\gamma/\nu \simeq 1.6$ and $z
\simeq 1.7$. For the $\lambda$-transition (Fig. 4) we obtain
$\beta \simeq 0.14$, $\nu \simeq 0.99$, $\gamma/\nu 
\simeq 1.6$ and $z \simeq 1.7$.
The comparison between the nonequilibrium static exponents
obtained here and the exact values corresponding to the
equilibrium Ising model is not straightforward. This is due
to several reasons: firstly, ours is a nonequilibrium model;
secondly, numerical errors involved are important; and thirdly,
one can expect crossover effects coming from the fact
that $\Delta x$ is not zero \cite{milchev86}. Concerning our
results for the dynamical exponent $z$, they are in accordance
with recent estimates for the equilibrium version of this model
\cite{mehlig92}. Furthermore, a simple dynamical scaling analysis
as the one performed in Ref. \cite{medina89} gives no changes of
the colored-noise exponents with respect to the
white-noise case. This is due to the fact that correlation of
the noise (\ref{eq:tdgl-cor}) is not a power law neither in space
nor in time. Hence in principle one cannot expect large changes in
our exponents as compared to the values of the equilibrium
$\phi^4$-model.

\section{An approximate theoretical approach}
\label{sec:spd}
The steady probability density of the nonmarkovian stochastic
process defined by Eq. (\ref{eq:tdgl}) can be found by
discretising space in a regular d-dimensional lattice
with spacing $\Delta x$. In this lattice, Eq. (\ref{eq:tdgl})
has the form (\ref{eq:dle}). The approximate Fokker-Planck
equation corresponding to this discrete Langevin equation
for small $\lambda$ and $\tau$ can be shown to be:
\begin{equation}
\label{eq:afpe}
\frac{\partial P}{\partial t} = \left(
-\frac{\partial}{\partial \psi_i}
f_i + \varepsilon \frac{\partial^2
}{\partial\psi_i^2}
+ \varepsilon \lambda^2 \nabla^2_{ik}\frac{\partial^2}
{\partial\psi_i \partial\psi_k}
+\varepsilon\tau  \frac{\partial^2}{\partial\psi_i
\partial\psi_k}\frac{\partial f_k}{\partial \psi_i}\right) P
\end{equation}
Details of the derivation of this equation are presented in
Appendix \ref{sec:appfpe}. The discretised version of the
laplacian operator can be written in terms of forward and
backward finite differences \cite{abramowitz}:
\begin{eqnarray}
\label{eq:lapl}
\nabla^2_{il}=\nabla^+_{ik}\nabla^-_{kl}\;,\;\;\rm{with}
\;\;\;\;\;\;\;\;
\nabla^+_{ij}\equiv \delta_{i+1,j}-\delta_{ij}\;,
\;\;
\nabla^-_{ij}\equiv \delta_{i,j}-\delta_{i-1,j}
\end{eqnarray}

It can easily be seen from these definitions that
$\nabla^+_{ij}=-\nabla^-_{ji}$. Moreover, the following
relations hold:
\begin{mathletters}
\label{eq:dr}
\begin{eqnarray}
\label{eq:dr1}
\frac{\partial}{\partial \psi_l}&\sum_j&
\left(\nabla^+ \psi\right)^2_j=
-2\left(\nabla^2\psi\right)_l  \\
\label{eq:dr2}
\frac{\partial}{\partial \psi_l}&\sum_j&\left(\nabla^2\psi
\right)^2_j=2 \left(\nabla^4\psi\right)_l
\\ \label{eq:dr3}
\frac{\partial}{\partial \psi_l}&\sum_j&\psi_j^3
\left(\nabla^2\psi\right)_j 
=\left(\nabla^2\psi^3\right)_l+3\psi^2_l
\left(\nabla^2\psi\right)_l
\\ \label{eq:dr4}
\frac{\partial}{\partial \psi_l}&\sum_j&\psi_j^2
\left(\nabla^2\psi^2\right)_j 
= 4\psi^2_l
\left(\nabla^2\psi\right)_l
\end{eqnarray}
\end{mathletters} 
The equation that the steady solution of (\ref{eq:afpe})
must verify is
\begin{equation}
\label{eq:sfpe}
\left(-f_i \; + \;\varepsilon \frac{\partial}{\partial\psi_i}
\;+\; \varepsilon \lambda^2 \nabla^2_{ik}
\frac{\partial}{\partial\psi_k}
\;+\;\varepsilon\tau  \frac{\partial}{\partial\psi_k}
\frac{\partial f_k}{\partial \psi_i}\right) P_{st}\;= \;0
\end{equation}
where the probability flux has been taken to be zero, as
usual. We will assume now that the solution of this equation
has the form \cite{schenzle85}
\begin{equation}
\label{eq:sol1}
P_{st}\sim e^{-{\displaystyle(F_0+F_1\tau+F_2\lambda^2)/
\varepsilon}}
\end{equation}
It is evident that $F_0$ corresponds to the solution of
the Fokker-Planck equation for the white-noise case ($\tau=0$,
$\lambda=0$):
\begin{equation}
\label{eq:wfpe}
\left(-f_i + \varepsilon \;\frac{\partial}{\partial\psi_i}
\right)P_{st}\;=\;0
\end{equation}
Introduction of $P_0\sim exp\left(-F_0/\varepsilon\right)$
leads to the following differential equation for $F_0$
\begin{equation}
\label{eq:def0}
\frac{\partial F_0}{\partial \psi_l} = - f_l
\end{equation}
which can be immediately solved after an inspection of
relation (\ref{eq:dr1}), leading to
\begin{equation}
\label{eq:sol2}
F_0=\frac{1}{4} \sum_k \left(-\psi_k^2+ \frac{1}{2} \psi_k^4
+\left(\nabla^+\psi\right)^2_k \right)
\end{equation}
The next step consists on introducing the ansatz
(\ref{eq:sol1}) into Eq. (\ref{eq:sfpe}). When only the first
nonzero orders in $\tau$ and $\lambda^2$ are considered and
Eq. (\ref{eq:def0}) is taken into account the following
equality is obtained:
\begin{eqnarray}
\label{eq:equality}
\left(-\tau \frac{\partial F_1}{\partial \psi_i}-\lambda^2
\frac{\partial F_2}{\partial \psi_i} + \lambda^2
 \nabla^2_{ik} f_k + \tau\varepsilon \frac
{\partial^2 f_k}{\partial \psi_k \partial \psi_i}
+\tau f_k \frac{\partial f_k}{\partial\psi_i}\right)P_{st}=0
\end{eqnarray}
And by comparing coefficients one finds the following
differential equations for $F_1$ and $F_2$:
\begin{equation}
\label{eq:f1}
\frac{\partial F_1}{\partial \psi_i} =
\varepsilon  \frac{\partial^2 f_k}{\partial \psi_k
\partial \psi_i} + 
f_k \frac{\partial f_k}{\partial \psi_i}
\end{equation}
\begin{equation}
\label{eq:f2}
\frac{\partial F_2}{\partial \psi_i}= \nabla^2_{ik} f_k
\end{equation}
These equations can be easily solved by considering the
definition of $f_k$ given in (\ref{eq:f4dfunc}) and taking
into account relations (\ref{eq:dr}). The solution is:
\begin{equation}
\label{eq:f1d}
F_1 = \frac{1}{4} \sum_i \left[ \frac{1}{2}
(1-12\varepsilon)\psi_i^2 - \psi_i^4 +\frac{1}{2}\psi_i^6 -
\left(\nabla^+ \psi \right)_i^2 - \psi_i^3\left(\nabla^2
\psi \right)_i +\frac{1}{2}\left(\nabla^2\psi\right)_i^2
\right]
\end{equation}
\begin{equation}
\label{eq:f2d}
F_2=\frac{1}{2}\sum_i \left[-\frac{1}{2}\left(\nabla^+
\psi \right)_i^2 + \frac{3}{4} \psi_i^2\left(\nabla^2\psi^2
\right)_i
-\psi_i^3\left(\nabla^2\psi \right)_i + \frac{1}{2}
\left(\nabla^2\psi\right)_i^2 \right]
\end{equation}
The discrete steady probability density is then found by
introducing (\ref{eq:sol2}), (\ref{eq:f1d}) and (\ref{eq:f2d})
into (\ref{eq:sol1}):
\begin{eqnarray}
\label{eq:stpdtau-1}
P_{st}\left(\vec{\psi}\right)&\sim&exp\left\{-\frac{1}{4
\varepsilon}
\sum_j\left[\left(-1+\frac{\tau}{2}-6\tau\varepsilon\right)
\psi_j^2+ \left(\frac{1}{2}-\tau\right)\psi_j^4
+\left(1-\tau-\lambda^2\right)\left(\nabla^+
\psi\right)^2_j
\right.\right.\nonumber \\&+&\left.\left.
\frac{\tau}{2}\psi_j^6
+\left(\lambda^2+\frac{\tau}{2}\right)\left(\nabla^2\psi
\right)^2_j-
\left(\tau+2\lambda^2\right)\psi_j^3\left(\nabla^2\psi
\right)_j+\frac{3}{2}\lambda^2\psi_j^2
\left(\nabla^2\psi^2 \right)_j\right]\right\}.
\end{eqnarray}
The last four terms of this expression can be shown to be irrelevant
by means of a renormalization-group
analysis \cite{wilson74}. Hence the expression for the
steady probability density in continuum space takes the form
\begin{eqnarray}
\label{eq:stpdtau-2}
P_{st}\sim exp\left\{-\frac{1}{4
\varepsilon}\int d\vec{x}\left[\left(-1+\frac{\tau}{2}
-6\tau\varepsilon\right) \psi^2+\left(\frac{1}{2}-
\tau\right)\psi^4+
\left(1-\tau-\lambda^2\right)\left(\nabla
\psi\right)^2
\right]\right\}.
\end{eqnarray}
This is the expression up to first order in $\tau$ and
$\lambda^2$ for the steady-state probability density of the
field variable $\psi$. It can be checked that in a mean-field
approach the result for a zero-dimensional nonmarkovian
stochastic process (i.e. $\lambda=0$ and $\tau \neq 0$)
is recovered \cite{sancho89}.
Since the effects of
$\tau$ and $\lambda$ are not clear from their appearance in
expression (\ref{eq:stpdtau-2}), we present in Appendix
\ref{sec:applin} an exactly solvable model in which one can
see the softening role of these two parameters.

\section{Comments and conclusions}
\label{sec:conc}

We have presented numerical and analytical results of the
effects of a non-white external noise in
the Time Dependent
Ginzburg-Landau model.
Since the system is close to the equilibrium case when the noise is
white, this model permits the comparison between a
well-known phenomenology of equilibrium and the phenomena we
expect in nonequilibrium situations. In this sense the noise
parameters $\lambda$ and $\tau$, which control the departure
from equilibrium, are assumed to be small in the theoretical
approximate approach.
A numerical analysis
of the model shows that the role of the correlation time and
length of the noise is to decrease its effective intensity.
This can be understood in a linear model which is solved
exactly in Appendix \ref{sec:applin}.

\begin{acknowledgements}
This research was supported in part by the Direcci\'on General de
Investigaci\'on Cient\'{\i}fica y T\'ecnica (Spain) under Project
No. PB90-0030. We acknowledge CESCA (Centre de
Supercomputaci\'o de Catalunya) and Cray Research, Inc.
for CPU time in their CRAYs YMP, where these calculations
have been performed. One of us (J.G.O.) thanks Fundaci\'o
Catalana per a la Recerca for financial support and the people
of the Chemistry Research and Engineering group at
Cray Research Park (Eagan, Minnesota) for their
hospitality and help while part of this work was carried out.
We also thank R. Toral for fruitful comments.

\appendix
\section{Fokker-Planck approximation for nonmarkovian
Langevin equations in extended systems}
\label{sec:appfpe}

Let us consider the discrete Langevin equation (\ref{eq:dle})
with a noise colored in space and time.
We are looking for an evolution equation
for the probability density
\begin{equation}
\label{eq:probdens}
P\left(\vec{\psi},t \right) = \left< \delta \left(
\vec{\psi}(t)
- \vec{\psi} \right) \right>
\end{equation}
where the average is taken over the initial conditions and
different realizations of the noise (this is the so-called
{\em Van Kampen's lemma} \cite{vankampen76}).
On the other hand, a continuity equation for the evolution of
$\left< \delta \left( \vec{\psi}(t)-\vec{\psi} \right)
\right>_{ic}$ (average taken over initial conditions only)
must hold:
\begin{equation}
\label{eq:sle}
\frac{\partial}{\partial t} \left< \delta \left(
\vec{\psi}(t)-
\vec{\psi} \right)\right>_{ic} = - 
\frac{\partial}{\partial\psi_i}
\;\dot{\psi}_i
 \left< \delta \left( \vec{\psi}(t)-\vec{\psi}
\right)\right>_{ic}
\end{equation}
This equation is the {\em stochastic Liouville equation}. Its
average over the noise $\xi_i(t)$ leads to an expression for
the evolution of the probability density defined above:
\begin{equation}
\label{eq:fpe1}
\frac{\partial P}{\partial t} = -
\frac{\partial}{\partial\psi_i}
f_i P - \frac{\partial}{\partial\psi_i} \left< \xi_i(t)
\delta \left( \vec{\psi}(t)-\vec{\psi} \right)\right>
\end{equation} 
The remaining average in (\ref{eq:fpe1}) can be calculated by
means of {\em Novikov's theorem} \cite{novikov65}:
\begin{equation}
\label{eq:novth}
\left< \xi_i(t)\delta \left( \vec{\psi}(t)-\vec{\psi}
\right)\right> = \int_0^t dt' \;\varepsilon\; h_{ij}(t,t')
\left< \frac{\delta\left( \delta \left( \vec{\psi}(t)-
\vec{\psi} \right)\right)}{\delta\xi_j(t')} \right>
\end{equation}
It can easily be seen that the following equality holds:
\begin{equation}
\label{eq:partdelta}
\frac{\delta\left(\delta \left( \vec{\psi}(t)-
\vec{\psi}\right)\right)}{\delta\xi_j(t')} =
-\frac{\partial}{\partial \psi_k}
\;\frac{\delta \psi_k(t)}{\delta \xi_j(t')}\; \delta \left( 
\vec{\psi}(t)-\vec{\psi} \right)
\end{equation}
This result, together with Novikov's theorem, leads to a new
expression for the equation we are trying to derive:
\begin{equation}
\label{eq:fpe2}
\frac{\partial P}{\partial t} = \frac{\partial}
{\partial\psi_i} f_i P + \frac{\partial^2}{\partial\psi_i
\partial \psi_k} \int_0^t dt'\; \varepsilon \;
h_{ij}(t,t') \left<\frac{\delta \psi_k(t)}{\delta \xi_j(t')}\;
\delta \left( \vec{\psi}(t)-\vec{\psi} \right)\right>
\end{equation}
The problem of the exact evaluation of this last average and
its relation to the probability density remains unsolved. Thus
an approximation has to be done at this point. Let us
assume that $\tau$ is small. Then the correlation function
$h_{ij}(t,t')$ of the noise appearing in the integral will be
a sharply peaked function of $t-t'$. This fact allows us to
use a Taylor-series expansion of the response function
appearing in the average of Eq. (\ref{eq:fpe2})
\begin{equation}
\label{eq:rptse}
\frac{\delta \psi_k(t)}{\delta \xi_j(t')} \simeq \left.
\frac{\delta \psi_k(t)}{\delta \xi_j(t')} \right |_{t=t'} +
\frac{d}{dt'} \left. \frac{\delta \psi_k(t)}{\delta \xi_j(t')}
\right|_{t=t'} (t'-t)
\end{equation}
We need to calculate now the response function at equal times
and its first time derivative. To do so we integrate formally
Eq. (\ref{eq:dle}) to obtain
\begin{equation}
\label{eq:fidle}
\psi_k(t)=\psi_k(0) + \int_0^t ds \; \left[f_k \left(
\vec{\psi}(s) \right) + \xi_k(s) \right]
\end{equation}
Functional differentiation of this expression leads through
(\ref{eq:rptse}) to
\cite{sancho89,hernandez83}
\begin{equation}
\frac{\delta \psi_k(t)}{\delta \xi_j(t')} \simeq
\delta_{kj}- \frac{\partial f_k\left(\vec{\psi}(t)\right)}
{\partial \psi_j(t)} (t'-t)
\end{equation}
By introducing this result into Eq. (\ref{eq:fpe2}), making use
of relation (\ref{eq:probdens}) and performing the integral in
time we find
\begin{equation}
\label{eq:fpe}
\frac{\partial P}{\partial t} = -\frac{\partial}{\partial
\psi_i} f_i P + \varepsilon \frac{\partial^2}{\partial\psi_i
\partial\psi_k} \left( h_{ik}^0+\tau  h_{ij}^1
\frac{\partial f_k}{\partial \psi_j} \right) P
\end{equation}
This is the final expression for the approximate Fokker-Planck
equation corresponding to a stochastic discretised field
process driven by an additive colored noise. Transient terms
have been neglected by extending the time integrals from $0$
to $\infty$, and also the following definitions have been used
\begin{equation}
h_{ij}^0 \equiv \int_0^{\infty} ds\; h_{ij}(s); \;\;\;\;\;
\tau h_{ij}^1 \equiv \int_0^{\infty} ds\; s\; h_{ij}(s)
\label{eq:h0h1def}
\end{equation}
These $h^k$ are functions in space with the same 
characteristic length $\lambda$. We assume that they are sharply peaked
around $\vec{x}-\vec{x}'= 0$, so that they can be worked out as
distributions having an expansion of the form
\begin{equation}
\label{eq:hexp}
h_{ik}^0=\delta_{ik}+ a_0\lambda^2\nabla^2_{im} \delta_{mk} ;
\;\;\;\; h_{ik}^1=a_1\delta_{ik}+ \theta(\lambda^2 )
\end{equation}
As the parameters $a_0$ and $a_1$ can be included in the
definition of $\tau, \lambda$ or $\varepsilon$, they can be
assumed to be equal to one. And as we want
to keep only the first corrections in $\tau$ and $\lambda^2$,
we can discard the dependence of $h^1$ on $\lambda^2$.
These expansions (\ref{eq:hexp}) can be understood in the following way:
the order 0 is the white-noise limit $\lambda\longrightarrow0$
(i.e. a Kronecker delta) and the first nonzero order
corresponds to $\lambda^2$ (due to space inversion symmetry),
which is supposed to be a laplacian operator (the simplest
spatial correlation beyond the Kronecker delta). Now,
introduction of (\ref{eq:hexp}) into Eq. (\ref{eq:fpe}) leads
finally to Eq. (\ref{eq:afpe}), which is the final approximate
Fokker-Planck equation up to the first non-zero orders in both
the correlation length and correlation time of the noise.

\section{Colored noise in a linear model}
\label{sec:applin}

In order to acquire a better understanding of the effect of
a colored noise in the steady state
behavior of our system let us consider a linear stable version of
Eq. (\ref{eq:tdgl}):
\begin{eqnarray}
\label{eq:ltdgl}
\frac{\partial\psi(\vec{x},t)}{\partial t} =
\frac{1}{2} \left( -\psi + 
\nabla^2 \psi \right) + \xi(\vec{x},t)
\end{eqnarray}
which in Fourier space takes the form:
\begin{equation}
\label{eq:ltdgl-f}
\frac{\partial\psi(\vec{q},t)}{\partial t} = -\frac{1}{2}
\left(1 + q^2\right) \psi(\vec{q},t) + \xi(\vec{q},t)
\end{equation}
The noise $\xi$ was defined in Eq. (\ref{eq:lambda}) and has a
correlation in Fourier space \cite{ojalvo92b}
\begin{equation}
\label{eq:cor-f}
\left< \;\xi(\vec{q},t) \;\xi(\vec{q}\,',t') \;\right> =
\frac{\varepsilon}{\tau (1+\lambda^2 q^2 )}\;\delta
(\vec{q}+\vec{q}\,') \; exp \left ( -( 1 + \lambda^2 q^2 ) \frac
{\mid t-t'\mid} {\tau} \right )
\end{equation}
The Fokker-Planck equation corresponding to the Langevin
equation (\ref{eq:ltdgl-f}) can be found by means of the
procedure described in Appendix \ref{sec:appfpe}. In this
case, however, as the Langevin equation is linear, the response
function can be evaluated {\em exactly} from Eq. (\ref{eq:fidle}).
The result is:
\begin{equation}
\frac{\delta\psi_q(t)}{\delta\xi_{q'}(t')} =
\delta_{qq'}\;e^{(t'-t)(1+q^2)/2}
\end{equation}
so that the (exact) Fokker-Planck equation is:
\begin{equation}
\label{eq:fpe-f}
\frac{\partial P}{\partial t} = \int d\vec{q}
\;\frac{\delta}{\delta \psi(\vec{q},t)} \frac{1}{2}
\left(1 + q^2\right) \psi(\vec{q},t) \;P + \int d\vec{q} \;
\varepsilon_{eff}(q) \;\frac{\delta}
{\delta\psi(\vec{q},t)} \;\frac{\delta P} {\delta\psi
(-\vec{q},t)}
\end{equation}
where transient terms have been neglected and the effective
noise intensity is given by
\begin{equation}
\label{eq:epseff}
\varepsilon_{eff}(q) = \frac{\varepsilon}{ (1+\lambda^2 q^2)
(1+\lambda^2 q^2 + \frac{\tau}{2}(1+q^2)) }
\end{equation}
The steady state probability distribution obeys
(assuming zero probability flux):
\begin{equation}
\label{eq:stfpe-f}
\left(\frac{1}{2}\left(1 + q^2 \right) \psi(\vec{q})\; +
\varepsilon_{eff}(q) \;\frac{\delta}{\delta\psi(-\vec{q})}
\right) P_{st}= 0
\end{equation}
which has the following solution
\begin{equation}
\label{eq:stprob2-f}
P_{st} \sim exp\left\{-\frac{1}{4} \int \frac{\left(1 + q^2
\right) \psi(\vec{q})\psi(-\vec{q})}{\varepsilon_{eff}(q)} \;
d\vec{q} \right\},
\end{equation}
as can be tested in a straightforward way by direct
substitution in (\ref{eq:stfpe-f}).
On the other hand, it can be easily seen from Eqs.
(\ref{eq:epseff}) and
(\ref{eq:stprob2-f}) that the steady probability density in
real space is: 
\begin{eqnarray}
\label{eq:stprob2}
P_{st} \sim exp\left\{-\frac{1}{4
\varepsilon}\int \left[\left(1+\frac{\tau}{2}
\right) \psi^2+
\left(1+\tau+\lambda^2\right)\left(\nabla
\psi\right)^2
+\left(\lambda^2+\frac{\tau}{2}\right)\left(\nabla^2\psi\right
)^2\right]d\vec{x}\right\}.
\end{eqnarray}
Thus we have found here that, in this exact
model, the role of the correlation time and length of the
noise is to decrease its effective intensity, as found in
our numerical simulation of the more general non-linear model.

\begin{figure}
\caption{
Spherically averaged spatial correlation function for the
noise driven by Eq. (\protect{\ref{eq:lambda}}) for four
different values of the
correlation length. The solid line corresponds to $\lambda=7$,
the dashed-dotted line to $\lambda=5$, the dashed line to
$\lambda=3$ and the dotted line to $\lambda=1$.
\label{fig:1}}
\end{figure}

\begin{figure}
\caption{
(a) Relative fluctuations of the order parameter versus the
intensity
of the noise for three values of the correlation time;
(b) phase diagram of the model.
Noise is correlated only in time ($\lambda = 0$).
\label{fig:2}}
\end{figure}

\begin{figure}
\caption{
Numerical results for $\varepsilon=0.7$ and $\lambda=0$.
The order parameter $M_1$ (a), the relative fluctuations of
the field $M_2$ (b) and the linear relaxation time $\tau_R$
(c) are shown against the inverse of the correlation time
of the noise
for different sizes of the system.
The broken lines are a guide to the eye.
In Fig. (a) empty stars correspond to an extrapolation to
infinite size. In Figs. (b) and (c) the vertical dashed lines
denote the position of the transition point also extrapolated
to the thermodynamic limit.
\label{fig:3}}
\end{figure}

\begin{figure}
\caption{
Numerical results for $\varepsilon=0.8$ and $\tau=0.3$.
The order parameter $M_1$ (a), the relative fluctuations of
the field $M_2$ (b) and the linear relaxation time $\tau_R$
(c) are shown against the inverse of the correlation length
of the noise for different sizes of the system.
The broken lines are a guide to the eye.
In Fig. (a) empty stars correspond to an extrapolation to
infinite size. In Figs. (b) and (c) the vertical dashed lines
denote the position of the transition point also extrapolated
to the thermodynamic limit.
\label{fig:4}}
\end{figure}

\end{document}